\documentclass[lettersize,journal]{IEEEtran}
\usepackage{amsmath,amsfonts}
\usepackage{algorithm,algorithmic}
\usepackage{array}
\usepackage{caption}
\usepackage[colorlinks,
            linkcolor=blue,       %%修改此处为你想要的颜色
            anchorcolor=blue,  %%修改此处为你想要的颜色
            citecolor=blue,        %%修改此处为你想要的颜色，例如修改blue为red
            ]{hyperref}
\usepackage[caption=false,font=normalsize,labelfont=sf,textfont=sf]{subfig}
\usepackage{textcomp}
\usepackage{stfloats}
\usepackage{url}
\usepackage{verbatim}
\usepackage{graphicx}
\usepackage{cite}
\usepackage{booktabs} %表格
\usepackage{balance}
\usepackage{multirow}
\usepackage[normalem]{ulem}
\usepackage{subfig}
\usepackage{xcolor}

\useunder{\uline}{\ul}{}
\def\BibTeX{{\rm B\kern-.05em{\sc i\kern-.025em b}\kern-.08em
    T\kern-.1667em\lower.7ex\hbox{E}\kern-.125emX}}
\begin{document}

\title{A Video Steganography for H.265/HEVC Based on Multiple CU Size and Block Structure Distortion}

\author{Xiang Zhang, Wen Jiang, Fei Peng, Wenbin Huang, Ziqiang Li, Zhangjie Fu

\thanks{This work was supported in part by the National Natural Science Foundation of China under Grant 62202234; in part by the China Postdoctoral Science Foundation under Grant 2023M741778. (\textit{Corresponding author: Zhangjie Fu, Wen Jiang}).

Xiang Zhang, Wen Jiang, Wenbin Huang, Ziqiang Li, and Zhangjie Fu are with the Engineering Research Center of Digital Forensics, Ministry of Education, Nanjing University of Information Science and Technology, Nanjing, Jiangsu 210044, China (e-mail: zhangxiang@nuist.edu.cn; 202312490517@nuist.edu.cn; wenbinhuang@nuist.edu.cn; iceli@mail.ustc.edu.cn; fzj@nuist.edu.cn).

Fei Peng is with the School of Artificial Intelligence, Guangzhou University, Guangzhou, Guangdong 510006, China (e-mail: eepengf@gmail.com).

%Min Long is with the School of Electronics and Communication Engineering, Guangzhou University, Guangzhou, Guangdong 510006, China (e-mail: %caslongm@aliyun.com).
}}
        % <-this % stops a space

\markboth{Journal of \LaTeX\ Class Files,~Vol.~14, No.~8, August~2021}%
{Shell \MakeLowercase{\textit{et al.}}: A Sample Article Using IEEEtran.cls for IEEE Journals}


\maketitle

\begin{abstract}
Video steganography based on block structure, which embeds secret information by modifying Coding Unit (CU) block structure of I-frames, is currently a research hotspot. However, the existing algorithms still suffer from the limitation of poor anti-steganalysis, which results from significantly disrupting the original CU block structure after embedding secret information. To overcome this limitation, this paper proposes a video steganography algorithm based on multiple CU size and block structure distortion. Our algorithm introduces three key innovations: 1) a CU Block Structure Stability Metric (CBSSM) based on CU block structure restoration phenomenon to reveal the reasons for the insufficient anti-steganalysis performance of current algorithms. 2) a novel mapping rule based on multiple CU size to reduce block structure change and enhance embedding capacity. 3) a three-level distortion function based on block structure to better guide the secret information embedding. This triple strategy ensures that the secret information embedding minimizes disruption to the original CU block structure while concealing it primarily in areas where block structure changes occur after recompression, ultimately enhancing the algorithm's anti-steganalysis. Comprehensive experimental results highlight the crucial role of the proposed CBSSM in evaluating anti-steganalysis performance even at a low embedding rate. Meanwhile, compared to State-of-the-Art video steganography algorithms based on block structure, our proposed steganography algorithm exhibits greater anti-steganalysis, as well as further improving visual quality, bitrate increase ratio and embedding capacity.
\end{abstract}

\begin{IEEEkeywords}
H.265/HEVC video steganography, Coding unit, Multiple CU size, Block structure distortion, CBSSM.
\end{IEEEkeywords}

\begin{figure}[!t]   % 使用 figure 而非 teaserfigure
    \centering
    \includegraphics[width=1\linewidth, trim=2cm 4cm 3cm 2cm, clip]{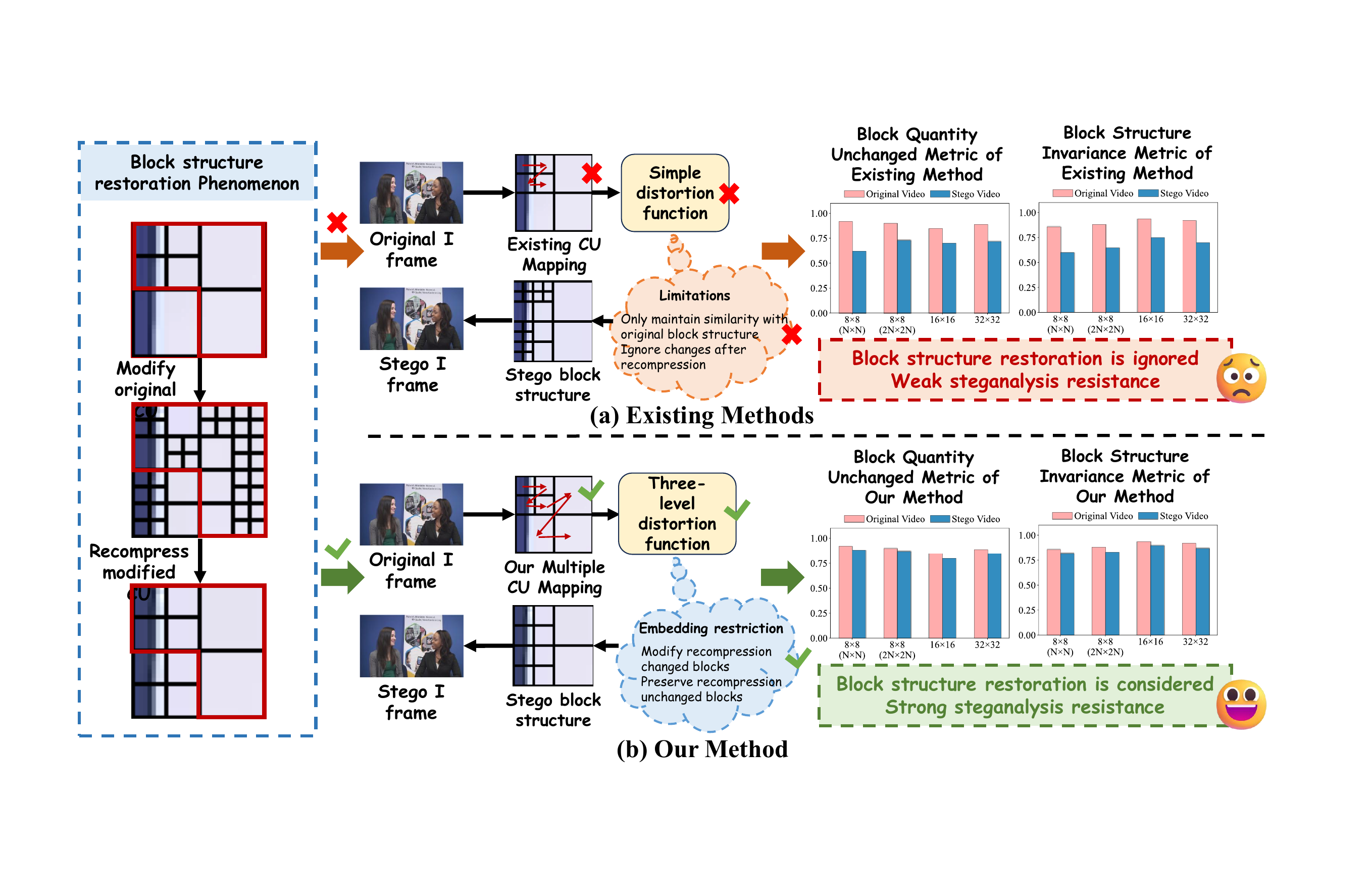}
    \caption{Comparison between existing methods and our proposed method: (a) Existing methods exhibit weak security when facing the block structure restoration phenomenon. (b) Our method designs a distortion function that considers the degree of block structure change before and after recompression, enhancing the security of the steganographic video.}
    \label{fig1}
\end{figure}

\section{Introduction}\label{sec1}
\IEEEPARstart{W}{ith} the development of information and multimedia technologies, ensuring the security of information transmission over public channels has become an urgent issue. Steganography is a technique that enables secure transmission of information by embedding secret data within multimedia files (such as images \cite{1}, audio \cite{2}, and video \cite{3}). Among these, video is one of the most significant types of multimedia files. Compared to images and audio, video contains much more redundant information, allowing it to carry a larger amount of secret data. Typically, video steganography requires integration with encoding technologies. H.265/HEVC is currently one of the widely used video encoding standards \cite{4}. Compared to the previous generation H.264/AVC, H.265/HEVC offers substantial advantages in compression efficiency, making it well-suited for high-resolution video applications. As a result, video steganography based on H.265/HEVC has become the mainstream approach in the field of video steganography.

Generally, video steganography based on H.265/HEVC can be categorized into three types based on different embedding carriers: prediction syntax elements, transform quantization coefficient, and block structure (I-frames). Among these, video steganography based on prediction syntax elements embeds secret information by modifying different syntax elements during the prediction process \cite{5,6,7,9,10,14,15,16,17}. Video steganography based on transform quantization coefficients embed secret information by altering the Discrete Cosine Transform (DCT) and Discrete Sine Transform (DST) coefficients \cite{18,19,20,21,22}. In recent years, researchers have developed video steganography based on block structure, which embeds information by modifying the CU block structure in I-frames \cite{23,24,25,26}, and offer low bitrate increase ratio and higher visual quality. However, they are easily detected by steganalysis. Their failure raises an intriguing question: \textit{\textbf{what is the fundamental difference between the stego video and the original video that can be exploited to design a universal steganalysis detection.}}

Video steganography based on block structure usually embeds secret information by modifying the CU block structure of I-frames. The existing scheme is shown in Fig.\ref{fig1}(a), these modifications have minimal impact on video quality, making the stego video appear almost identical to the original from the pixel perspective. However, we have observed an intriguing phenomenon that reveals significant differences between stego and original video from the block structure perspective. We refer to this intriguing phenomenon as \textit{\textbf{CU block structure restoration}: regardless of how the CU block structure is modified in H.265/HEVC, it tends to restore to a form similar to the original after recompression.} Based on the phenomenon, we propose a CU Block Structure Stability Metric (CBSSM), which contains Block Quantity Unchanged Metric (BQUM) and Block Structure Invariance Metric (BSIM). By calculating block quantity unchanged rate and block structure invariance rate after recompression, this metric can quantitatively analyze the block structure difference between the stego and original video. 
 
According to above analysis, we can conclude that keeping the block structure unchanged and embedding secret information in the CU whose block structure is not restored are the core of resisting steganalysis. Therefore, we propose a new mapping rule and a three-level distortion, which is shown in Fig.\ref{fig1}(b). Specifically, the mapping rule maps multiple CU size in I-frames into binary sequence. Then, by taking the binary sequence as the carriers, the secret information is hidden by modifying the structure of each CU in no more than one depth, effectively preserving the block structure with minimal changes after steganography while enhancing embedding capacity. Meanwhile, we design a three-level distortion function based on block structure, the distortion assigns cost based on the extent of CU block structure changes after recompression. Higher cost is assigned to the block with minimal structure change, while lower cost is given to that with significant changes. This forces the secret information to be embedded in CUs whose block structures are altered significantly (not restored) after recompression, thereby improving the anti-steganalysis performance. To sum up, our contributions are as follows:
\begin{itemize}
\item \textbf{A CU Block Structure Restoration Phenomenon}. We are the first to provide theoretical insight to explain an interesting CU block structure restoration phenomenon in H.265/HEVC based on the RDO difference bound and the Lipschitz assumption.
\item \textbf{A Mapping Rule Based on Multiple CU Size}. Based on the phenomenon, we propose a new mapping rule which maps multiple CU size in I-frames into binary sequence.
\item \textbf{A Three-level Distortion Function Based on Block Structure}. By analyzing the CU block structure change before and after recompression, we design a new three-level distortion function to guide information hiding.
\item \textbf{Performance Evaluation with Different Comparison algorithms}. We conduct extensive experiments and demonstrate the better performance in various aspects compared to State-of-the-Art algorithms.
\end{itemize}

\section{Related Work}\label{sec2}
In the field of video steganography based on block structure, Tew \textit{et al.} \cite{23} proposed the first steganography algorithm based on different CU size. It embeds secret bits by modifying CUs with different sizes into $8\times8$ CUs. This algorithm has good visual quality and low bitrate increase ratio. Dong \textit{et al}. \cite{24} proposed a Steganographic Compression Efficiency Degradation Model (SCEDM), which computes the Kullback–Leibler (KL) divergence between the stego CU block structure and the original CU block structure to quantify the distributional disparity between them. Among four mapping strategies, the one yielding the smallest KL divergence is selected for embedding secret information. This model effectively enhances the visual quality of the stego video. Yang \textit{et al.} \cite{25} introduced Split Flag-based Cover Mapping (SFCM) method based on CU size, which first maps the CU block structure to a sequence. Afterwards, they proposed Maintenance Principle of Quad-tree Structures (MPQS) to transform the distortion minimization of cover sequence into similarity between the depth vectors of the stego and cover. The secret information is embedded according to MPQS, which further reduce the distortion caused by steganography. Wang \textit{et al.} \cite{26} proposed an adaptive video steganography algorithm. This algorithm first maps secret information to different block structures in $8\times8$ CUs. Then, the Rate Distortion Optimization (RDO) is introduced to establish an adaptive distortion function for STC\cite{27}. This algorithm has higher embedding capacity and visual quality. Although the steganography algorithms based on block structure described above can obtain good visual quality, they disrupt the CU block structure of original videos during the embedding process, which resulting in poor resistance to steganalysis. Therefore, we propose a video steganography based on multiple CU size and block structure distortion to solve the insufficient anti-steganalysis performance.

\section{proposed Method}\label{sec3}
\begin{figure}[htbp]
    \centering
    \includegraphics[width=\linewidth, trim=0cm 0cm 0cm 0cm, clip]{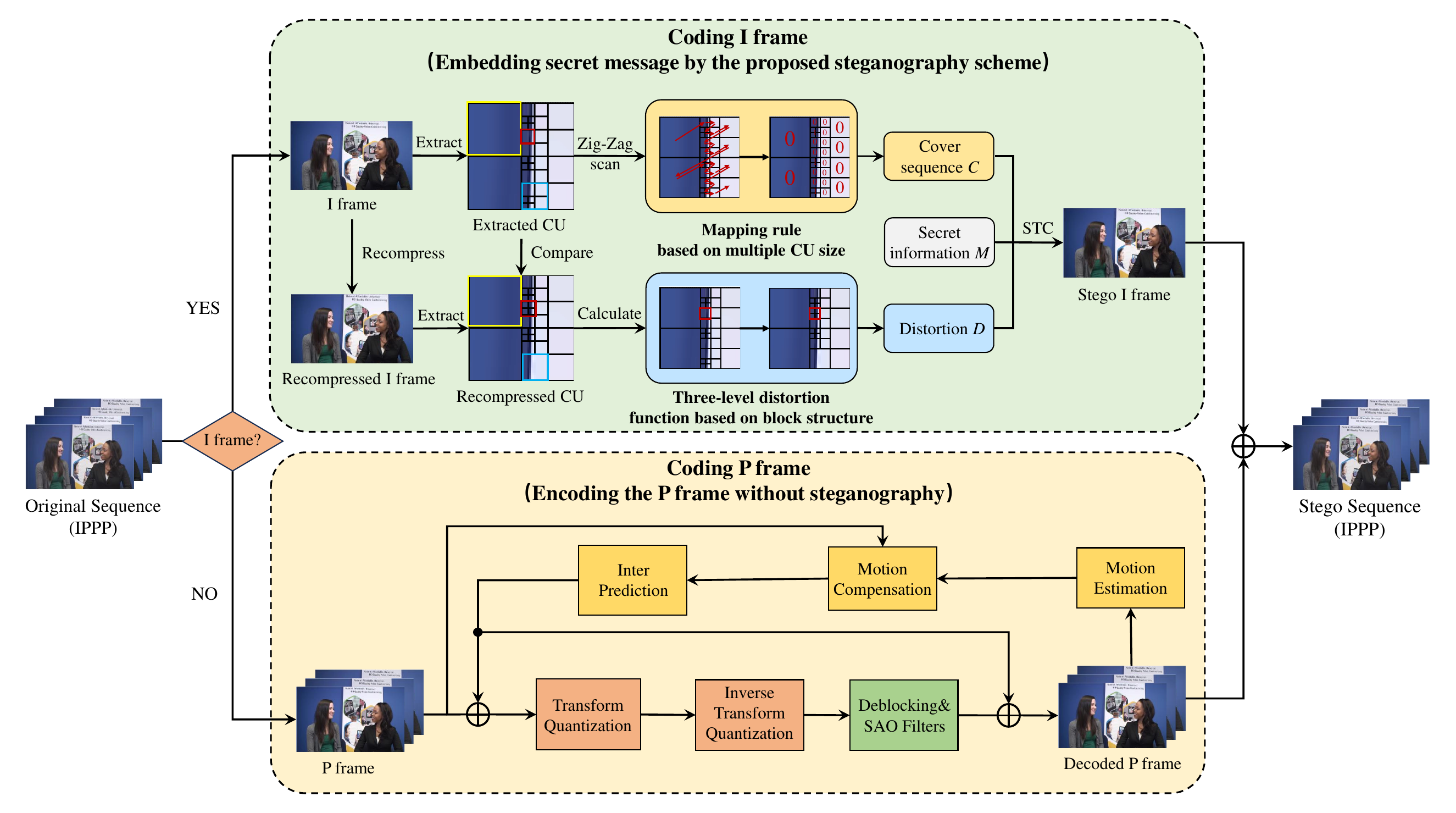}
    \caption{The framework of the proposed scheme}
    \label{Fig.3}
\end{figure}

In this section, we introduce the complete framework of the proposed video steganography scheme which is illustrated in Fig.\ref{Fig.3}. We use I-frames and P-frames to encode the original video. Taking the GOP “IPPP”, if the current encoded frame is a P-frame, no steganography is performed. If the current encoded frame is an I-frame, we first extract each CU in the I-frame. Meanwhile, we recompress the I-frame and then extract the recompressed CU to calculate the three-level distortion. Finally, the proposed mapping rule based on multiple CU size and the three-level distortion function are used to embed secret information into the I-frame.

\subsection{CU Block Structure Restoration Phenomenon}\label{sec3a}
\begin{figure}[h]
    \centering
    \includegraphics[width=1\linewidth, trim=0cm 4cm 0cm 4cm, clip]{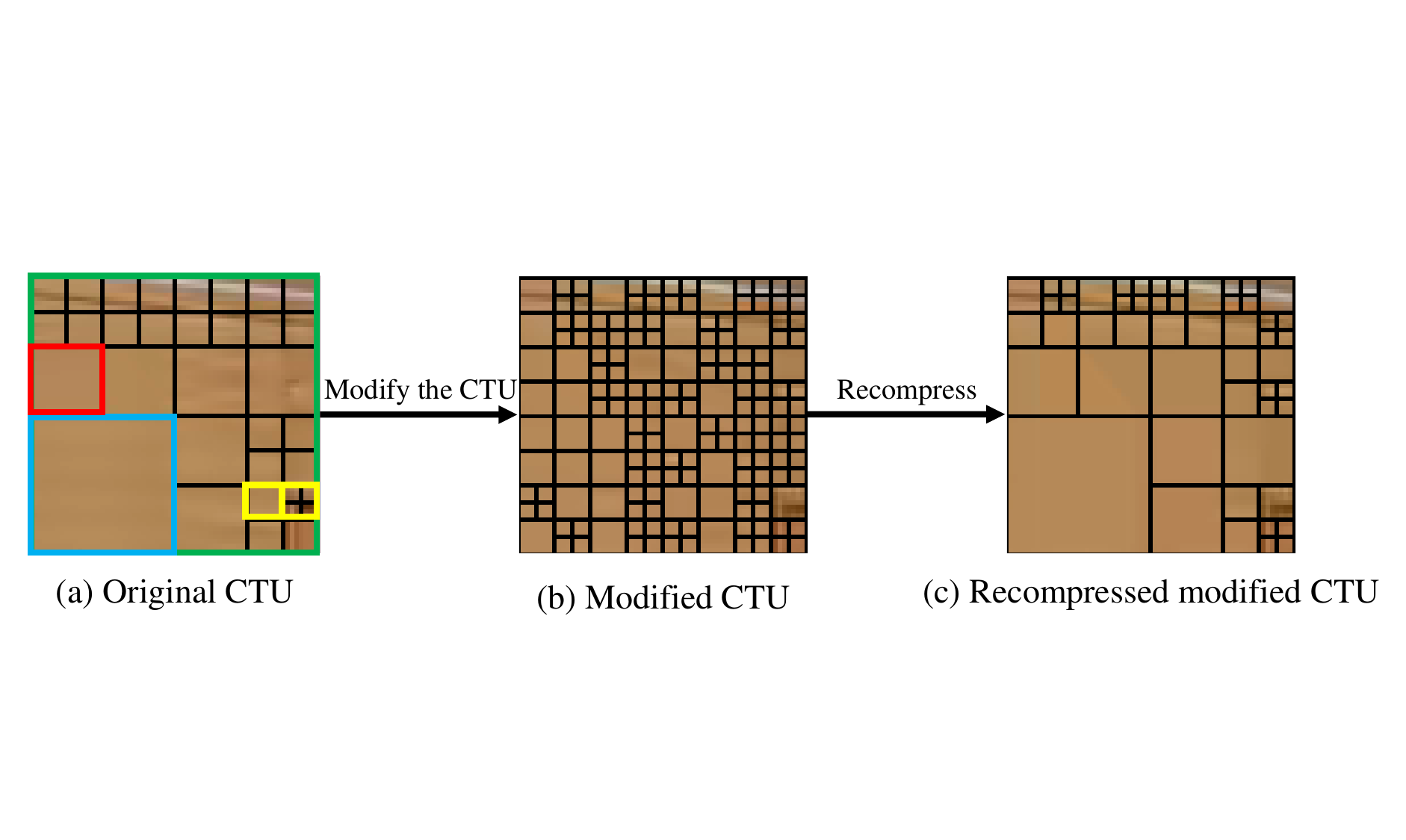}    %左下右上
    \caption{The block structure restoration phenomenon}
    \label{Fig.4}
\end{figure}

In order to better illustrate the CU block structure restoration phenomenon intuitively, we have done an experiment using ``Basketballpass" sequence with $QP$=32, profile ``encoder\_lowdelay\_P\_main", and the Group of Pictures (GOP) ``IPPP". The experiment results are shown in Fig. \ref{Fig.4}. 

The original example CTU, shown in Fig. \ref{Fig.4}(a) is obtained from the \#1 frame of ``Basketballpass" sequence. Then, we significantly and randomly modify the original CTU's block structure to generate the modified CTU, shown in Fig. \ref{Fig.4}(b). Finally, we recompress the video and extract the same CTU, as shown in Fig. \ref{Fig.4}(c). Comparing the three sub-figures, we observe that despite the serious modifications to the block structure of original CTU, the recompressed block structure remains highly similar to the original one, demonstrating the block structure restoration phenomenon. To further explain this intriguing phenomenon, we theoretically analyzed the fundamental reason for CU block structure restoration phenomenon as follows:

Let the pixel of a CU be denoted as $x$. During the initial encoding process, the candidate CU block structure set for this CU is represented by $\mathcal{CS}$. For each block structure $cs \in \mathcal{CS}$, denote the distortion and rate under this block structure as $D_{cs}(x)$ and $R_{cs}(x)$, respectively. The RDO cost function is defined as:
\begin{equation}
J_{cs}(x) = D_{cs}(x) + \lambda R_{cs}(x),
\label{eq:rdo_cost}
\end{equation}
where $\lambda > 0$ is the Lagrange multiplier determined by quantization parameter. During recompression, the input pixel becomes $x' = x + \delta$, where $\delta$ represents the perturbation caused by quantization. The perturbation magnitude satisfies $\|\delta\| \le \varepsilon$, where $\|\cdot\|$ denotes either the Euclidean norm.

\textbf{Lipschitz Assumption:} There exist constants $L_J \ge 0$, for any block structure $cs$ and perturbation $\|\delta\| \le \varepsilon$, we have:
\begin{equation}
- L_J\|\delta\| \le J_{cs}(x+\delta) - J_{cs}(x) \le L_J\|\delta\|.
\label{eq:bound2}
\end{equation}

As the predictor of H.265/HEVC typically exhibits a locally linear or interpolated form, distortion is computed as the sum of squared errors, and bitrate estimation depends smoothly on the residual distribution and entropy model. Therefore, this assumption is reasonable.

\textbf{Objective:} Let the optimal block structure before recompression be ${cs}^{\star} = \arg\min_{{cs} \in \mathcal{CS}} J_{cs}(x).$ Define the cost margin between the optimal block structure and any other block structure as:
\begin{equation}
\Delta_{cs} = J_{cs}(x) - J_{{cs}^{\star}}(x).
\label{eq:margin}
\end{equation}

We aim to prove that if $\Delta_{cs} > B(\varepsilon)$, for some lower bound $B(\varepsilon)$ determined by the perturbation strength $\varepsilon$, then the optimal block structure remains unchanged after recompression.

\textbf{Theorem 1: Sufficient Condition for Block Structure Restoration}\label{tm1}: If for all ${cs} \neq {cs}^{\star}$,
\begin{equation}
\Delta_{cs} = J_{cs}(x) - J_{{cs}^{\star}}(x) > 2L_J\varepsilon,
\label{eq:sufficient}
\end{equation}
then, for any perturbation $\|\delta\| \le \varepsilon$, the optimal block structure after recompression remains $cs^{\star}$.

\textbf{Proof:} For any ${cs} \neq {cs}^{\star}$, consider the cost difference after recompression:
\begin{equation}
\begin{aligned}
J_{cs}(x+\delta) - J_{{cs}^{\star}}(x+\delta)
= &\big[J_{cs}(x+\delta) - J_{cs}(x)\big]  \\
&+ \big[J_{cs}(x) - J_{{cs}^{\star}}(x)\big] \\
\quad &+ \big[J_{{cs}^{\star}}(x) - J_{{cs}^{\star}}(x+\delta)\big].
\end{aligned}
\label{eq:proof1}
\end{equation}
Applying the bound in~\eqref{eq:bound2}, we obtain:
\begin{equation}
\begin{aligned}
J_{cs}(x+\delta) - J_{{cs}^{\star}}(x+\delta)
\ge -L_J\|\delta\| + \Delta_{cs} - L_J\|\delta\|&
\\= \Delta_{cs} - 2L_J\|\delta\|.
\end{aligned}
\label{eq:proof2}
\end{equation}

If $\Delta_{cs} > 2L_J\varepsilon$ and $\|\delta\| \le \varepsilon$, then $J_{cs}(x+\delta) - J_{{cs}^{\star}}(x+\delta) > 0$,  indicating that ${cs}^{\star}$ remains the optimal block structure. Therefore, The RDO bound of $cs$ change is $B(\varepsilon) = 2L_J\varepsilon$. 

\begin{figure}[htbp]
    \centering
\includegraphics[width=0.7\linewidth]{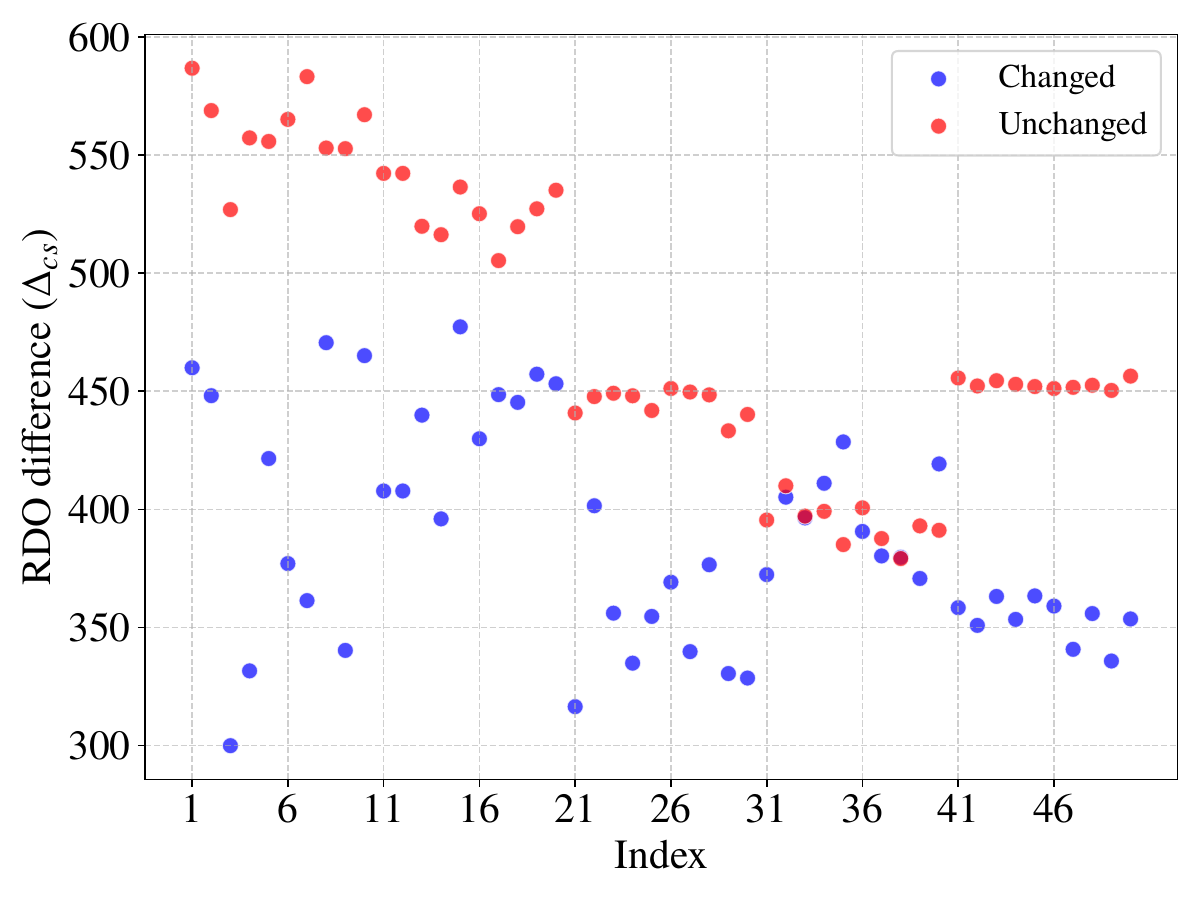}
\caption{The average RDO difference distribution map.}
\label{fig_RDO_diff1}
\end{figure}

The existence of this bound can also be verified experimentally. To simplify the analysis, we focus on identifying whether a distinct RDO difference bound exists between the suboptimal and optimal block structure, five representative video sequences with varying texture complexity and motion characteristics are selected, with resolutions ranging from  $416\times240$ to $2560\times1600$: ``BasketballPass", ``BasketballDrill", ``FourPeople", ``BasketballDrive", and ``PeopleOnStreet". We encode the first 10 frames in each video, for a total of 50 frames, and the RDO difference ($\Delta_{cs}$) of all CUs are calculated for each frame. The sequences are then classified two categories based on whether their block structures changed after recompression and the average $\Delta_{cs}$ of the two categories in each frame are illustrated in Fig. \ref{fig_RDO_diff1}. 

In the figure, the horizontal axis indicates the indices of the 50 selected frames, and the vertical axis shows the average $\Delta_{cs}$ for each frame, blue dots represent the average $\Delta_{cs}$ of CUs whose block structures changed in these frames, while red dots correspond to those that remained unchanged. As observed, a distinct bound exists between these two groups, and the RDO difference for blocks whose structures remain unchanged after recompression (red dots) is significantly larger than that for blocks whose structures are altered (blue dots) in most frames, further validating the rationality of the proposed block structure restoration theory.

To approximate the bound value $B(\varepsilon)$, the $\Delta_{cs}$ of each CU is normalized by its $\varepsilon$ value according to our statistical results, and the value of $L_J$ is approximately 1.1, implying that the bound value is around $2.2\varepsilon$. This suggests that when the $\Delta_{cs}$ of a CU exceeds $2.2\varepsilon$, its block structure usually remains unchanged after recompression. Due to the intrinsic RDO mechanism in H.265/HEVC, Our statistical results show that more than 85\% of the CUs satisfy this condition on average.

\subsection{CU Block Structure Stability Metric}\label{sec3b}
Based on the above findings, we designed a CU Block Structure Stability Metric (CBSSM) which contains Block Quantity Unchanged Metric ($BQUM$) and Block Structure Invariance Metric ($BSIM$) to reveal the limitation of current methods in resisting steganalysis. As can be seen from Section \ref{sec3a}, There are four sizes of CUs: $64\times64$ (green box in Fig. \ref{Fig.4}(a)), $32\times32$ (blue box in Fig. \ref{Fig.4}(a)), $16\times16$ (red box in Fig. \ref{Fig.4}(a)), and $8\times8$ (yellow box in Fig. \ref{Fig.4}(a)). As shown in Fig. \ref{Fig.4}(a), CUs of size $8\times8$ exhibit two distinct block structures: $8\times8$ without division, shown as the left yellow box in Fig. \ref{Fig.4}(a) (refer as $8\times8, 2N \times 2N$ in the following) and $8\times8$ divided into four $4\times4$ sub-blocks, shown as the right yellow box in Fig. \ref{Fig.4}(a)(refer as $8\times8, N \times N$ in the following). Since modifying a $64\times64$ CU introduces noticeable steganography traces, it is generally avoided as a carrier in video steganography. Therefore, the block structure we discuss here is represented as a set $T= \{(32 \times 32), (16 \times 16),  (8 \times 8, 2N \times 2N), (8 \times 8, N \times N)\}$. Based on the four block structures in $T$, we first define the block quantity unchanged metric $BQUM_t^i$ for $t$ ($t\in T$) block structure in the $i^{th}$ frame, which is shown as:
\begin{equation}
BQUM_t^i=\frac{1}{\mathrm{exp}(\frac{\left|N_t^i-\overline {N}_{t}^{i}\right|}{N_t^i})}, t\in T \text{,}
\end{equation}
where $N_t^i$ represents the CU number of the $t$ block structure in the $i^{th}$ frame of the original video, $\overline {N}_{t}^{i}$ is that of the recompressed video, $\mathrm{exp}(\cdot)$ is an exponential function used to normalize $BQUM_t^i$ to the range between 0 and 1. $BQUM_t^i$ is a ratio that quantifies the block quantity unchanged of the $t$ block structure in the $i^{th}$ frame before and after recompression. A lower value $BQUM_t^i$ indicates a greater change in block quantity, whereas a higher value means a minimal change. Since most video steganography algorithms based on block structure significantly modifies the CU block structure of the original frame, and structure restoration phenomenon will occur after recompression. This results in a substantial variation in block quantity of each block structure. Consequently, $BQUM_t^i$ decreases significantly.

Next, we further define the block structure invariance metric $BSIM^i_t$ for the $t$ block structure in the $i^{th}$ frame after recompression, which is shown as:
\begin{equation}
BSIM^i_t=\frac{\sum_{k=1}^{N_t^i}\rho(cs_t^i(k),\overline {cs}_t^{i}(k))}{N_t^i},t\in T \text{,}
\end{equation}
where $cs_t^i(k)$ is the block structure of the CU in $t$ block structure at $k$ position with Zig-Zag scan in the $i^{th}$ frame of the original video ($cs_t^i(k)=t$). while $\overline {cs}_t^{i}(k)$ is that of the recompressed CU. $\rho(\cdot)$ is a contrast function, as defined by:
\begin{equation}
\rho(cs_t^i(k),\overline {cs}_t^{i}(k))=
\begin{cases}
1,\quad cs_t^i(k)=\overline {cs}_t^{i}(k) \\
0,\quad cs_t^i(k) \neq \overline {cs}_t^{i}(k),
\end{cases} t\in T.
\end{equation}

 $BSIM^i_t$ is a ratio that quantifies the block structure invariance of the $t$ block structure in the $i^{th}$ frame before and after recompression. A lower value $BSIM^i_t$ indicates a greater change in block structure, whereas a higher value suggests a minimal change. Video steganography based on block structure significantly modifies the block structure of the original frame, resulting in a significant decline in $BSIM_t^i$.
 
  In order to further verify the effectiveness of our proposed CBSSM, we conduct experiments on $BQUM$ and $BSIM$. Firstly, we compress five video sequences, including ``Traffic'', ``PeopleOnStreet'', ``BQTerrace'', ``BasketballDrive'', ``FourPeople'', using $QP$ = 32. We then decode and recompress these videos, calculating the $BQUM$ and $BSIM$ between the recompressed videos and the original videos. 
 
 \begin{figure}[htbp]
  \centering
  \scalebox{0.9}{%
    \begin{minipage}{\linewidth}
      \captionsetup[subfigure]{labelfont={scriptsize}, textfont={scriptsize}, justification=centering}
      \subfloat[\scriptsize Tew's \cite{23} Block Quantity Unchanged Metric (BQUM)]{%
        \includegraphics[width=0.48\linewidth]{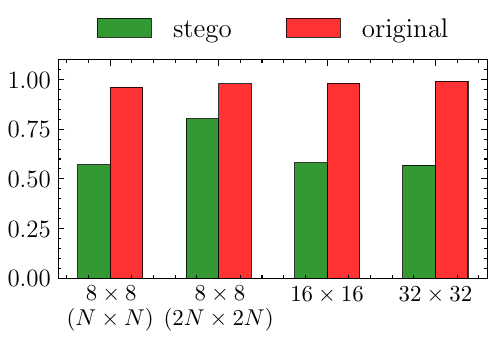}
      }
      \hfill
      \subfloat[\scriptsize Tew's \cite{23} Block Structure Invariance Metric (BSIM)]{%
        \includegraphics[width=0.48\linewidth]{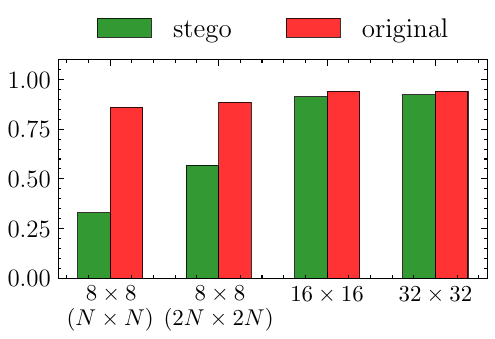}
      }
      \caption{The average CBSSM features of five video sequences for detecting Tew \cite{23}}
      \label{img.CBSSM}
    \end{minipage}
  }
\end{figure}
 
  The average results for the first 10 I-frames of the five video sequences are represented by the red bars in Fig. \ref{img.CBSSM}. Secondly, video steganography algorithm based on block structure, Tew \cite{23} is used to embed the secret information into these videos, with a payload of 0.5 bits per cover (bpc). After generating the stego videos, we evaluate the $BQUM$ and $BSIM$ between the recompressed stego videos and the original stego videos. The average results for the first 10 I-frames of the five video sequences are represented by the green bars in Fig. \ref{img.CBSSM}. The figure clearly shows a significant reduction in both $BQUM$ and $BSIM$ after applying the steganography algorithm, which preliminarily demonstrates that the existing video steganography algorithms based on block structure can be easily detected by the proposed $BQUM$ and $BSIM$, as it disrupts the CU block structure.

 \subsection{Mapping Rule Based on Multiple CU Size}\label{sec3c}
To preserve the original CU structure as much as possible during the steganography process, we restrict the modification depth of the CU to no more than one. This approach ensures that the CU size of the recompressed stego video closely resembles that of the original stego video. To achieve the above objectives, we first design a mapping rule to map multiple CU sizes into a binary sequence. Specifically, assume that the size of all the CUs except for $64 \times 64$ in the $i^{th}$ I-frame by Zig-Zag scan as $CS^i=\{cs^i(1),cs^i(2),...,cs^i(q)\}$, where $q$ means the total number of CU in the $i^{th}$ I-frame. We exclude $64\times64$ blocks since they usually in the smooth areas, and embedding information in these areas could noticeably degrade quality and increase bitrate. Then, we map CU sizes $CS^i$ into binary sequence $C^i=\{c^i(1),c^i(2),...,c^i(q)\}$. The $j^{th}$ element in the binary sequences $C^i$ is obtained by the mapping rule as follows: 
\begin{equation}\label{eq15}
c^i(j)=\left\{\begin{array}{ll}0,
& \mathrm{if}\,\,\,\, cs^i(j)=32 \times 32 \\0, 
& \mathrm{if}\,\,\,\, cs^i(j)=16 \times 16 \\0, 
& \mathrm{if}\,\,\,\, cs^i(j) =8 \times 8, \,2N \times 2N\\1, 
& \mathrm{if}\,\,\,\, cs^i(j)=8 \times 8, \,N \times N,
\end{array}\right.
\end{equation}
where $c^i(j)$ represent the current element of the binary sequence $C^i$. Subsequently, the binary sequence $C^i$ is treated as the carrier which is input into the STC algorithm with the secret information $M$ and three-level distortion function $D$ described in Section \ref{sec3d} for steganography. The detailed steganography process is presented in Section \ref{sec3e}.

\subsection{Three-level Distortion Function Based on Block Structure}\label{sec3d}
\begin{figure}[htpb]
    \centering
    \includegraphics[width=0.7\linewidth, trim=1cm 0.5cm 1.1cm 1cm, clip]{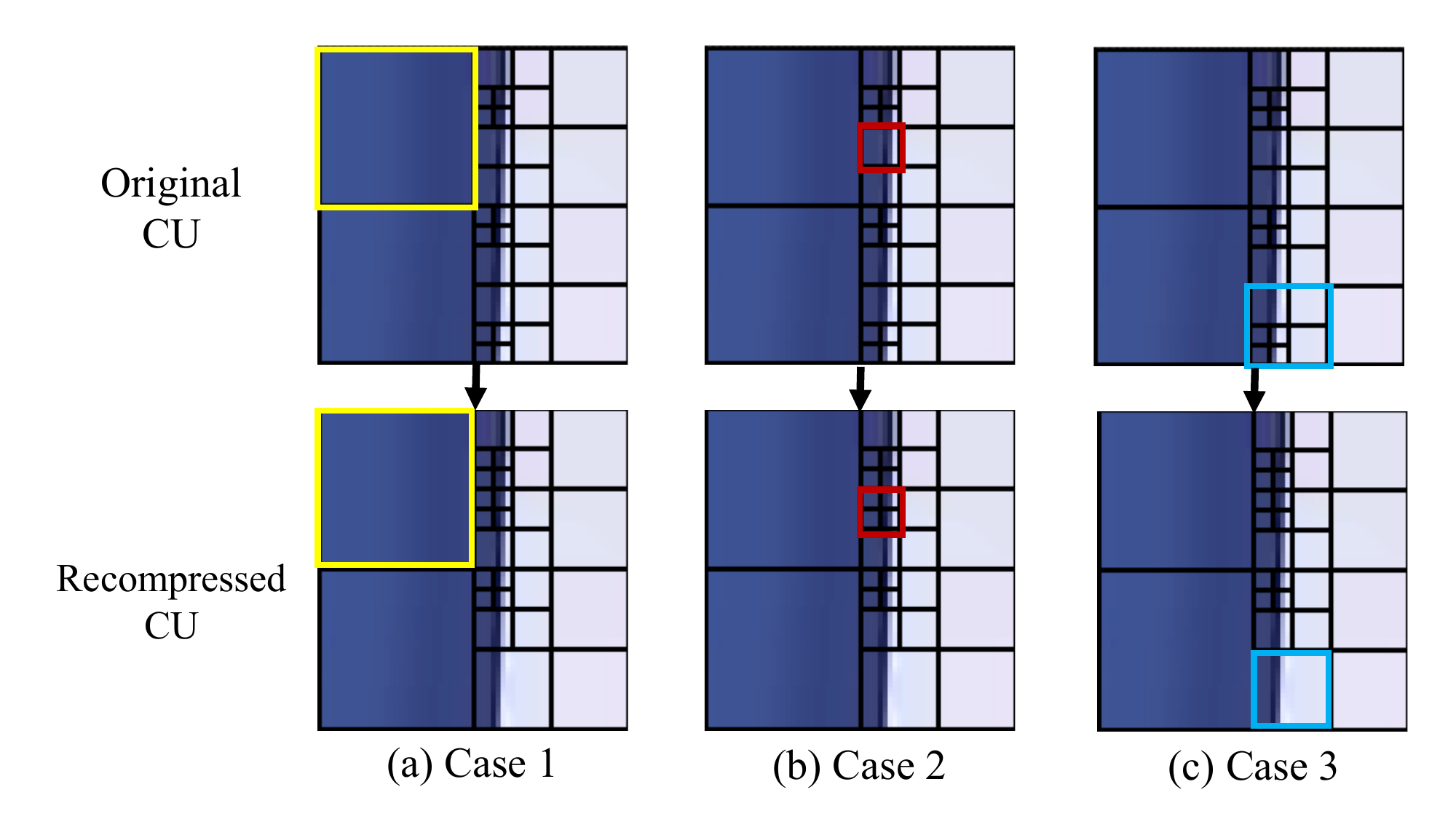}    %左下右上
    \caption{Example CTUs for all cases}
    \label{fig5}
\end{figure}

As discussed in Section \ref{sec3b}, CUs whose block structure change after recompression are suitable for modification in steganography. Therefore, it is necessary to design a distortion function that allocates different costs to different CUs. The allocation rule is as follows: CUs with block structures that change after recompression are assigned lower costs, while those that remain unchanged receive higher costs. Based on this, we develop a three-level distortion function based on block structure with RD value $J$ whose calculation method can refer to Equation (\ref{eq:rdo_cost}). We first define the maximum depth of a CU as:
\begin{equation}\label{eq17}
\begin{aligned}
MD(CU^i(k))= \left\{\begin{array}{ll}1,
 \,\,\mathrm{if}\,\,\,\,SB(CU^i(k))=32 \times 32 \\2, 
 \,\,\mathrm{if}\,\,\,\,SB(CU^i(k))=16 \times 16 \\3, 
 \,\,\mathrm{if}\,\,\,\,SB(CU^i(k))=8 \times 8, \,2N \times 2N\\4, 
 \,\,\mathrm{if}\,\,\,\,SB(CU^i(k))=8 \times 8, \,N \times N,
\end{array}\right.
\end{aligned}
\end{equation}
where $CU^i(k)$ represents the CU at position $k$ with Zig-Zag scan in the $i^{th}$ frame of the original video, $MD(CU^i(k))$ represents maximum depth of $CU^i(k)$, $SB(CU^i(k))$ represents the smallest block size in $CU^i(k)$. For example, the smallest block size of the CU marked in blue in Fig. \ref{fig5}(c) is $8 \times 8, \,N \times N$, and its maximum depth is four. Afterwards, we recompress $CU^i(k)$ to obtain $\overline {CU}^i(k)$, and calculate the maximum depth difference $MDD$ between the $CU^i(k)$ and $\overline {CU}^i(k)$ as:
\begin{equation}
MDD(CU^i(k))=|MD(CU^i(k))-MD(\overline {CU}^i(k)) |,
\end{equation}

Based on the value of $MDD(CU^i(k))$, we classify $CU^i(k)$ into the following three cases: 

Case 1: if $MDD(CU^i(k))=0$, it indicates that the block structure of the original $CU^i(k)$ is identical to that of the recompressed $\overline {CU}^i(k)$, making it unsuitable for steganography. Therefore, a larger distortion needs to be assigned to it. The example  CU of this case is shown in Fig. \ref{fig5}(a) and marked by yellow.

Case 2: if $MDD(CU^i(k))=1$, this indicates that a block structure change with a maximum depth of one has occurred between the original $CU^i(k)$ and the recompressed $\overline {CU}^i(k)$. As a result, this CU is suitable for steganography, requiring a lower distortion allocation. The example CU of this case is shown in Fig. \ref{fig5}(b) and marked by red.

Case 3: if $MDD(CU^i(k))>1$, this means that there is a dramatic change between the block structure of the original $CU^i(k)$  and the recompressed $\overline {CU}^i(k)$. Therefore, $CU^i(k)$ is very suitable for steganography, and we need to allocate it the lowest distortion. The example CU of this case is shown in Fig. \ref{fig5}(c) and marked by blue.  

As can be seen in Fig. \ref{fig5}(c), the original minimum block in the blue box is $8\times8$ ($N\times N$), corresponding to a maximum depth of 4. After recompression, as shown in Fig. \ref{fig5}(c), the minimum block becomes $16\times16$, corresponding to a maximum depth of 2. Therefore, $MDD(CU^i(k))$ is equal to 2, which meets the condition of being greater than 1.

We then design the three-level distortion function $D(CU^i(k))$ for the above three cases as follows:
\begin{equation}\label{eq18}
D(CU^i(k))=
\begin{cases}
MD(CU^i(k))\cdot DR(CU^i(k)),  &\mathrm{if\,\,\,\, Case\,1} \\
DR(CU^i(k)),  &\mathrm{if \,\,\,\,Case\,2} \\
\frac{1}{MDD(CU^i(k))}\cdot DR(CU^i(k)),  &\mathrm{if\,\,\,\, Case\,3},
\end{cases}\quad
\end{equation}
where $DR(\cdot)$ represents RD difference rate calculation function and defined as:
\begin{equation}
DR(CU^i(k))=\frac{|J(CU^i(k))-J(CU^{'i}(k))|}{J(CU^i(k))},
\end{equation}
where $CU^{'i}(k)$ represents the $CU^i(k)$ after steganography. From Equation (\ref{eq18}), we observe that for Case 1, the distortion is greater than $DR(CU^i(k))$, and as the maximum depth $MD(CU^i(k))$ increases, the distortion also increases. For Case 2, the distortion is equal to $DR(CU^i(k))$. while in Case 3, the distortion is less than $DR(CU^i(k))$, and as the maximum depth difference $MDD(CU^i(k))$ increases, the smaller the distortion.

\subsection{Steganography Scheme Based on Multiple CU Size and Three-level distortion}\label{sec3e}
In this section, we introduce the specific process of the proposed steganography scheme. Firstly, the original video stream is decoded to extract the size of all the CUs $CS^i$ except for $64 \times 64$ in the $i^{th}$ I-frame by Zig-Zag scan as:
\begin{equation}
CS^i=\{cs^i(1),cs^i(2),...,cs^i(q)\}.
\end{equation}

Secondly, according to Equation (\ref{eq15}), we map $CS^i$ into binary sequence as:
\begin{equation}
C^i=\{c^i(1),c^i(2),...,c^i(q)\}.
\end{equation}

We then calculate the three-level distortion function $D(CU^i(k))$ for each CU in the $i^{th}$ I-frame by Zig-Zag scan according to Equation (\ref{eq18}).

Thirdly, generate the binary secret information $M$ by a pseudo-random function. Select a payload of $\alpha$. Input $C^i$, $M$, and $\alpha$ into STC algorithm $STC(\cdot)$ with the distortion function $D$ to obtain the stego binary sequence $S^i$ as:
\begin{equation}
S^i=STC(C^i,M,\alpha,D).
\end{equation}

Finally, in order to ensure that the modification depth of the CU is within one, we designed a block structure modification rule to complete the final steganography combined with the mapping rule. Assume that $c^i(j)$ and $s^i(j)$ are the current element of the binary sequence and stego binary sequence, respectively. The specific modification rule is as follows. 

\textbf{Rule 1}: if $c^i(j)=s^i(j)$, the block structure of the current CU remains unchanged.

\textbf{Rule 2}: if $c^i(j)=0\,\,\,$and$\,\,\,s^i(j)=1$, the current CU is divided into four sub-blocks.

\textbf{Rule 3}: if $c^i(j)=1\,\,\,$and$\,\,\,s^i(j)=0$, the current CU is merged into $8 \times 8, \,\,2N \times 2N$.

It is worth noting that in order to increase capacity, the secret information bit 0 corresponds to three different block sizes, if the block structure cannot be uniquely identified, errors may occur at the extraction end. To avoid this issue, we present two solutions: (1) by using the same version and configuration of the H.265/HEVC encoder and compress the identical video sequence at the extraction stage, the original block structure can be obtained, and compare it with block structure of stego video to obtain the secret binary sequence. (2) transmit the original block structure information as auxiliary side information.

Furthermore, the two solutions mentioned above may not be so convenient to deploy in practice. Considering the needs of practical deployment, we also propose a variant of the algorithm to address the extraction issues. We only use $8\times8$ CUs as carriers for secret information. Then, these $8\times8$ CUs are mapped to binary sequences $C^i=\{c^i(1),c^i(2),...,c^i(q)\}$. The $j^{th}$ element in the binary sequences $C^i$ is obtained by the prposed mapping rule (only $8\times8$) as follows:

\begin{equation}\label{eq:variant}
c^i(j)=\left\{\begin{array}{ll}0, 
& \mathrm{if}\,\,\,\, cs^i(j) =8 \times 8, \,2N \times 2N\\1, 
& \mathrm{if}\,\,\,\, cs^i(j)=8 \times 8, \,N \times N,
\end{array}\right.
\end{equation}

This variant scheme only modifies $8\times8$ CUs. Therefore, it is able to successfully extract secret information without requiring any other information.

\section{Experimental Results and Analysis}\label{sec6}

\subsection{Experimental Setup}\label{sec6a}
All experiments are carried out using the H.265/HEVC reference software HM 16.9, the computer configuration is Intel (R) Core (TM) i9-12900 KF, 3.1 GHz, 32 GB memory. A total of 32 standard H.265/HEVC test sequences are shown in TABLE \ref{tab1} as our test videos, with resolution ranging from $416\times240$ to $2560\times1600$, using the profile ``encoder\_lowdelay\_P\_main'', and the GOP is ``IPPP''. The $QP$ is set to 26, 32 and 38, respectively. For each $QP$, three different payloads 0.1 bpc, 0.3 bpc, and 0.5 bpc are tested. The first 100 frames of each video sequence are used for steganography. Our proposed video steganography algorithm is compared with the four State-of-the Art steganography algorithms based on block structure, namely, Tew \cite{23}, Dong \cite{24}, Yang\cite{25}, Wang \cite{26}, where Proposed denotes the proposed scheme that modifies all block sizes other than $64\times64$, and Proposed ($8\times8$) denotes the proposed variant scheme that modifies only $8\times8$ blocks.

\begin{table}[h]
    \centering
        \caption{\small YUV TEST SEQUENCE}
    \label{tab1}
    \resizebox{\linewidth}{!}{
    \scriptsize
    % 2. 减小列间距（默认 6pt，改为 3pt 或更小）
    \setlength{\tabcolsep}{3pt}
    % 3. 减小行间距（默认 1，改为 0.8 或 0.7）
    \renewcommand{\arraystretch}{0.9}
    \begin{tabular}{lll}
            \hline
    \hline
        CLASS & Resolution & Sequence\\
        \hline
        CLASS\_A & \(2560\times1600\) & PeopleOnStreet, Traffic, NebutaFestival \\
        \hline
        CLASS\_B & \(1920\times1080\) & BasketballDrive, BQTerrace, Cactus,\\ 
        & &ParkScene, Kimonol, Tennis,\\ 
        & &blue\_sky,  crowd\_run \\
        \hline
        CLASS\_C & \(832\times480\) & BasketballDrill, BasketballDrillText,\\ 
        & &BQMall, PartyScene, RaceHorsesC,\\ 
        & &Flowervase, Mobisode2  \\
        \hline
       \multirow{2}{*}{CLASS\_D} & \multirow{2}{*}{\(416\times240\)} & BasketballPass, BlowingBubbles,\\
         & &  BQSquare, RaceHorses, Keiba  \\
        \hline
        CLASS\_E & \(1280\times720\) & FourPeople, Johnny, KristenAndSara,\\
        & &SlideEditing, SlideShow, mobcal\_ter, \\
        & &vidyo1, vidyo3, vidyo4\\
        \hline
        \hline
    \end{tabular}
    }
\end{table}

\subsection{Analysis of Subjective Visual Quality }\label{sec6b}
\begin{figure}[htpb]
    \centering
    \includegraphics[width=0.8\linewidth, trim=0cm 0cm 0cm 0cm, clip]{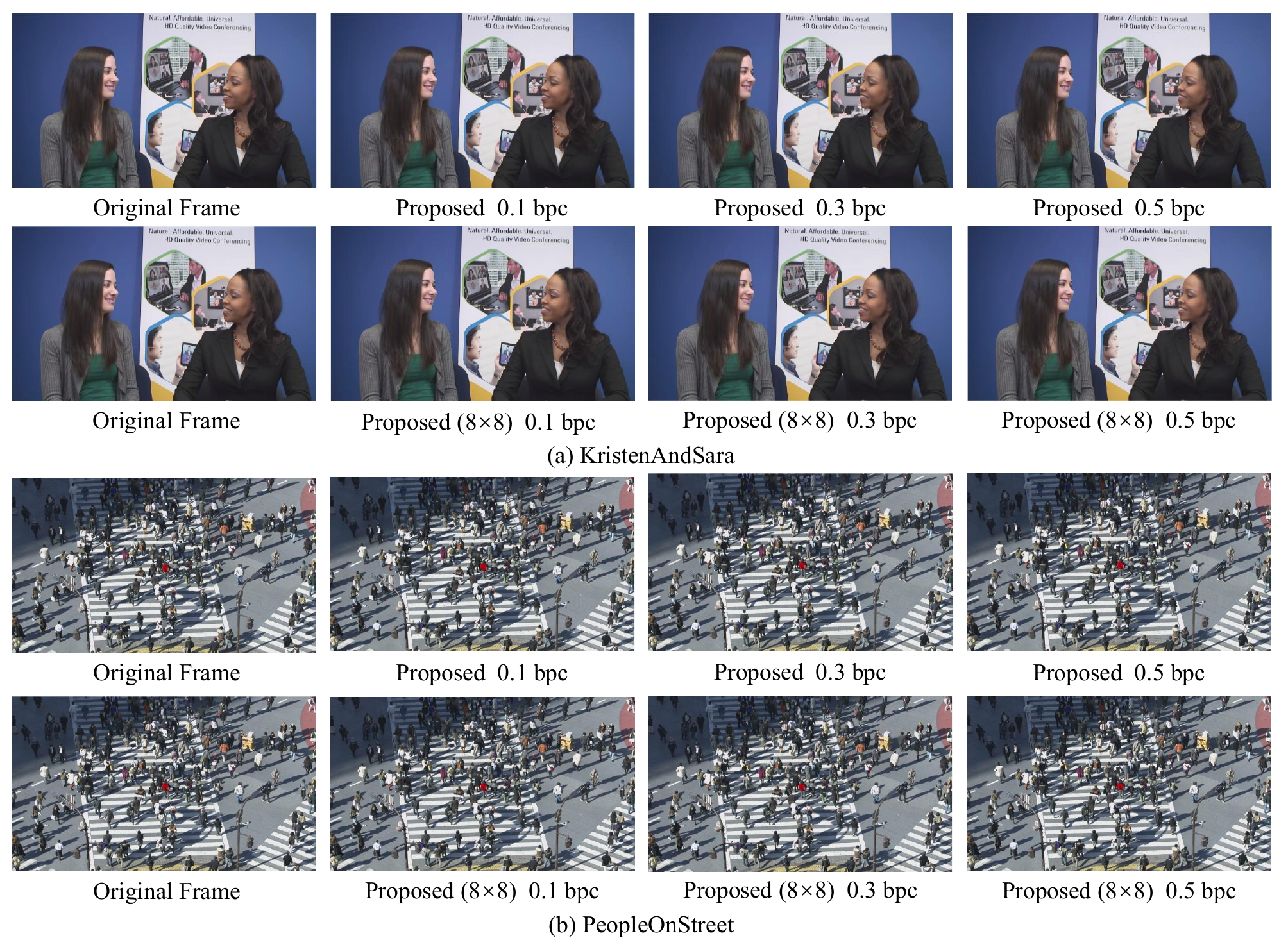}
    \caption{Comparison of subjective visual quality with different payload}
    \label{fig9}
\end{figure}

An effective video steganography algorithm should make it impossible for human eye to distinguish between the original and the stego videos. Since the texture of ``KristenAndSara'' is simple while that of ``PeopleOnStreet'' is relatively complex. We choose the two sequences as the example for subjective visual quality analysis. Fig. \ref{fig9} shows the \#1 frame in original video and stego videos with $QP=26$, payload =  0.1 bpc, 0.3 bpc, and 0.5 bpc, respectively. From Fig \ref {fig9}, it can be seen that at three different payloads, we cannot find any difference between the original frames the stego frames from human eyes. Consequently, both our proposed schemes have good subjective visual quality.

\subsection{Analysis of Objective Visual Quality}\label{sec6c}
In this section, Peak Signal-to-Noise Ratio ($PSNR$) is used to measure the objective visual quality of the proposed algorithm and the four compared algorithms, to be fair comparison, we follow the testing method proposed by Wang \cite{26} and calculate the $\Delta PSNR$ caused by embedding 1000 bits secret information, the $\Delta PSNR$ defined as:
\begin{equation}
    \Delta PSNR=1000 \cdot \frac{|PSNR_{ori}-PSNR_{stg}|}{capacity},
\end{equation}
where $PSNR_{ori}$ and $PSNR_{stg}$ represent the $PSNR$ value of the original video and the stego video. $capacity$ is the total bits of secret information embedded in the video. 

The average $\Delta PSNR$ of different $QPs$ and payloads of all six schemes are shown in TABLE \ref{tab2}, where the optimal results are marked in bold and the suboptimal results are underlined. The results in the following tables of the other indicators are also highlighted in the same way. In the table, the symbol $\downarrow$ represents that a lower $\Delta PSNR$ value is better. As shown in the table, the two schemes we proposed achieve the optimal and suboptimal $\Delta PSNR$, respectively. The optimal $\Delta PSNR$ achieves an average improvement of about 34\% compared to Wang’s result \cite{26}. This improvement can be attributed to the proposed mapping rule, which ensures that each block is modified only once, preventing any block from undergoing skip-level modifications. Meanwhile, our proposed distortion function also considers the RDO value, thereby enhancing the objective visual quality. The $\Delta PSNR$ of Proposed ($8\times8$) is not as good as that of Proposed because the capacity of modifying only $8\times8$ blocks is smaller than that of Proposed, resulting in a slightly larger $\Delta PSNR$ value for Proposed ($8\times8$). While Wang \cite{26} is better than Tew \cite{23}, Dong\cite{24} and Yang \cite{25}, because it does not significantly disrupt the optimal block structure by only modifying the $8\times8$ CUs. However, the other three methods cause more significant changes to the block structure, which is contrary to the RDO process, leading to visual quality reduction. 

\begin{table}[htpb]
\caption{The average $\Delta PSNR$ (\textnormal{dB} $\cdot10^{-2}\downarrow$) of the six schemes}\label{tab2}
\resizebox{\linewidth}{!}{
\scriptsize
% 2. 减小列间距（默认 6pt，改为 3pt 或更小）
\setlength{\tabcolsep}{3pt}
% 3. 减小行间距（默认 1，改为 0.8 或 0.7）
\renewcommand{\arraystretch}{0.9}
\begin{tabular}{cccccccc}
\hline
\hline
\(QP\)              & Sequence & Tew\cite{23}          & Dong\cite{24}   & Yang\cite{25}   & Wang\cite{26}         & Proposed ($8\times8$) & Proposed        \\ \hline
\multirow{5}{*}{26} & CLASS\_A & 0.0199       & 0.0490 & 0.0313 & 0.0082       & {\ul 0.0079}                       & \textbf{0.0067} \\
                    & CLASS\_B & 0.0372       & 0.1135 & 0.0674 & {\ul 0.0209} & 0.0210                             & \textbf{0.0158} \\
                    & CLASS\_C & 0.2739       & 0.8616 & 0.5029 & 0.2005       & {\ul 0.1951}                       & \textbf{0.1201} \\
                    & CLASS\_D & 0.6579       & 1.1866 & 0.8815 & 0.3513       & {\ul 0.2808}                       & \textbf{0.1771} \\
                    & CLASS\_E & 0.1713       & 0.4770 & 0.2558 & 0.1046       & {\ul 0.1021}                       & \textbf{0.0638} \\ \hline
\multirow{5}{*}{32} & CLASS\_A & 0.0262       & 0.0790 & 0.0415 & 0.0146       & \textbf{0.0102}                    & {\ul 0.0140}    \\
                    & CLASS\_B & 0.0556       & 0.2358 & 0.1405 & 0.0433       & {\ul 0.0410}                       & \textbf{0.0369} \\
                    & CLASS\_C & 0.4436       & 1.6344 & 1.0876 & 0.4121       & {\ul 0.3974}                       & \textbf{0.2810} \\
                    & CLASS\_D & 0.8679       & 2.1242 & 2.0964 & 0.5330       & \textbf{0.4388}                    & {\ul 0.4587}    \\
                    & CLASS\_E & 0.2881       & 0.9655 & 0.7339 & 0.2273       & {\ul 0.2134}                       & \textbf{0.1752} \\ \hline
\multirow{5}{*}{38} & CLASS\_A & 0.0405       & 0.1465 & 0.0729 & 0.0323       & {\ul 0.0300}                       & \textbf{0.0200} \\
                    & CLASS\_B & {\ul0.0811}       & 0.4206 & 0.2481 & 0.0966       &  0.0918                       & \textbf{0.0618} \\
                    & CLASS\_C & {\ul 0.6034} & 3.0299 & 2.1058 & 0.9966       & 0.8376                             & \textbf{0.5024} \\
                    & CLASS\_D & 1.3609       & 4.0059 & 2.4373 & 0.6964       & {\ul 0.6255}                       & \textbf{0.5438} \\
                    & CLASS\_E & {\ul 0.3697} & 1.5770 & 0.9586 & 0.4582       & 0.4390                             & \textbf{0.2754} \\ \hline
\multicolumn{2}{c}{Average}    & 0.3531       & 1.1271 & 0.7774 & 0.2797       & {\ul 0.2488}                       & \textbf{0.1835} \\ 
\hline
\hline
\end{tabular}
}
\end{table}

\subsection{Analysis of Bitrate Change}\label{sec6d}
In this section, Bitrate Increase Ratio ($BIR$) is used to evaluate the impact of the six steganography algorithms on video bitrate. Similarly we calculate the $BIR$ value caused by embedding 1000 bit secret information. The definition of $BIR$ is provided as:
\begin{equation}
    BIR=1000 \cdot \frac{|Bit_{stg}-Bit_{ori}|}{capacity \cdot Bit_{ori}},
\end{equation}
where $Bit_{stg}$ is the bitrate of the compressed video after steganography, $Bit_{ori}$ is the bitrate of the original compressed video. TABLE \ref{tab3} presents the average $BIR$ results for the six compared steganography algorithms. In the table, the symbol $\downarrow$ represents that a lower $BIR$ value is better.

\begin{table}[htpb]
\caption{The average $BIR$ $(10^{-2} \downarrow)$ of the six schemes}\label{tab3} % 表格标题
\resizebox{\linewidth}{!}{
\scriptsize
% 2. 减小列间距（默认 6pt，改为 3pt 或更小）
\setlength{\tabcolsep}{3pt}
% 3. 减小行间距（默认 1，改为 0.8 或 0.7）
\renewcommand{\arraystretch}{0.9}
\begin{tabular}{cccccccc}
\hline
\hline
\(QP\)              & Sequence & Tew\cite{23}             & Dong\cite{24}   & Yang\cite{25}   & Wang\cite{26}            & Proposed ($8\times8$) & Proposed        \\ \hline
\multirow{5}{*}{26} & CLASS\_A & 0.0069          & 0.0147 & 0.0095 & {\ul 0.0046}    & \textbf{0.0045}                    & 0.0047          \\
                    & CLASS\_B & 0.0189          & 0.0491 & 0.0328 & {\ul 0.0141}    & \textbf{0.0139}                    & 0.0148          \\
                    & CLASS\_C & \textbf{0.1104} & 0.3453 & 0.2409 & 0.1587          & 0.1345                             & {\ul 0.1200}    \\
                    & CLASS\_D & 0.2287          & 0.3746 & 0.3689 & \textbf{0.1024} & {\ul 0.1032}                       & 0.1109          \\
                    & CLASS\_E & {\ul 0.0642}    & 0.2077 & 0.1361 & 0.0669          & 0.0667                             & \textbf{0.0638} \\ \hline
\multirow{5}{*}{32} & CLASS\_A & 0.0121          & 0.0253 & 0.0149 & {\ul 0.0098}    & 0.0100                             & \textbf{0.0084} \\
                    & CLASS\_B & 0.0362          & 0.0833 & 0.0545 & 0.0282          & {\ul 0.0276}                       & \textbf{0.0261} \\
                    & CLASS\_C & 0.3040          & 0.5841 & 0.4397 & 0.2751          & {\ul 0.2717}                       & \textbf{0.2487} \\
                    & CLASS\_D & 0.3235          & 0.6487 & 0.5315 & {\ul 0.1908}    & 0.1974                             & \textbf{0.1877} \\
                    & CLASS\_E & 0.1396          & 0.3153 & 0.2022 & 0.1214          & {\ul 0.1199}                       & \textbf{0.1074} \\ \hline
\multirow{5}{*}{38} & CLASS\_A & 0.0214          & 0.0434 & 0.0245 & 0.0211          & {\ul 0.0210}                       & \textbf{0.0171} \\
                    & CLASS\_B & 0.0637          & 0.1330 & 0.0937 & 0.0620          & {\ul 0.0614}                       & \textbf{0.0493} \\
                    & CLASS\_C & {\ul 0.6012}    & 1.1121 & 0.9301 & 0.6525          & 0.6824                             & \textbf{0.5070} \\
                    & CLASS\_D & 0.5817          & 1.1379 & 0.6909 & 0.3955          & \textbf{0.3681}                    & {\ul 0.3875}    \\
                    & CLASS\_E & 0.2469          & 0.4933 & 0.3367 & {\ul 0.2192}    & 0.2215                             & \textbf{0.1886} \\ \hline
\multicolumn{2}{c}{Average}    & 0.1840          & 0.3712 & 0.2738 & 0.1548          & {\ul 0.1536}                       & \textbf{0.1361} \\ 
\hline
\hline
\end{tabular}
}
\end{table}

From TABLE \ref{tab3}, it can be observed that the $BIR$ of Proposed is the lowest among the six compared schemes on most test videos. This is because Proposed modifies the block structure only once during the embedding process, reducing the bitrate increase. Three-level distortion function selectively embeds information into CUs with the lowest distortion cost, further reducing the bitrate increase. Moreover, restricting modifications to only $8\times8$ blocks eliminates the flexibility to select larger, more efficient CUs during RDO process, resulting in a higher embedding distortion and a slightly larger $BIR$ value, yet it still ranks suboptimal among the compared methods. While for Tew \cite{23}, CUs with different sizes are forcibly converted into $8\times8$ CUs, which results in a significant change to the block structure. As a consequence, the bitrate increases rapidly. For Dong \cite{24}, they use the SCEDM model to control the distribution differences in the types and quantities of stego CUs. However, it does not maintain structure similarity of CUs, which disrupts the original RDO process, thereby increasing the bitrate. For Yang \cite{25}, SFCM mapping rule and the MPQS principle are employed to ensure the block structure remains as similar as possible after steganography. However, as the embedding capacity increases and multi-stage STC embedding is applied, many $32\times32$ CUs have been modified, leading to bitrate growth. For Wang \cite{26} modifying only the $8\times8$ CUs for steganography can maintain a low bitrate increments.

\subsection{Analysis of Capacity}\label{sec6e}
In our experiment, the capacity results are standardized using the $BIR$ and represent the number of bits embedded when the bitrate changes by 1\%. The average capacity results are shown in TABLE \ref{tab4}. In the table, the symbol $\uparrow$ represents that a larger capacity is better. As seen in the table, Proposed achieves the highest capacity across most test videos, with an average of 50,972 bits per 1\% bitrate increase. This can be attributed to the mapping rules and the three-level distortion function designed in Proposed. They both effectively suppressing the growth of $BIR$. Consequently, under the same $BIR$ variation conditions, Proposed achieves a higher embedding capacity. Since the embedding carrier of Proposed ($8\times8$) is limited to only $8\times8$ blocks, the number of available carriers for steganography is reduced, leading to a decrease in embedding capacity. However, its average capacity still ranks second. The embedding capacity of Wang\cite{26} is lower than that of Proposed ($8\times8$) because its bitrate increase is higher. The capacity of Tew\cite{23}, Dong\cite{24}, and Yang\cite{25} are relative low because they significantly disrupt the CU block structure. 
%In summary, the capacities of the Tew\cite{23}, %Dong\cite{24}, Yang\cite{25} and Wang\cite{26} are %only 78.6\%, 32.5\%, 48.8\% and 95.9\% of our %Proposed algorithm on average, respectively. 

\begin{table}[htpb]
\caption{The average capacity (\textnormal{bits} $\uparrow$) of the six schemes}\label{tab4}
\resizebox{\linewidth}{!}{
\scriptsize
% 2. 减小列间距（默认 6pt，改为 3pt 或更小）
\setlength{\tabcolsep}{3pt}
% 3. 减小行间距（默认 1，改为 0.8 或 0.7）
\renewcommand{\arraystretch}{0.9}
\begin{tabular}{cccccccc}
\hline
\hline
\(QP\)              & Sequence & Tew\cite{23}         & Dong\cite{24}  & Yang\cite{25}   & Wang\cite{26}          & Proposed ($8\times8$) & Proposed        \\ \hline
\multirow{5}{*}{26} & CLASS\_A & 255991      & 85221 & 111415 & {\ul 269685}  & 262629                             & \textbf{274347} \\
                    & CLASS\_B & 75765       & 31925 & 52777  & {\ul 130768}  & \textbf{131874}                    & 113225          \\
                    & CLASS\_C & 11432       & 5123  & 7221   & {\ul 15150}   & \textbf{15983}                     & 15142           \\
                    & CLASS\_D & 4474        & 3000  & 2806   & {\ul 11217}   & \textbf{11250}                     & 10096           \\
                    & CLASS\_E & {\ul 15681} & 4863  & 7369   & 15082         & 15123                              & \textbf{15762}  \\ \hline
\multirow{5}{*}{32} & CLASS\_A & 103938      & 52699 & 74626  & 110710        & {\ul 116523}                       & \textbf{145479} \\
                    & CLASS\_B & 35568       & 17017 & 32233  & {\ul 64206}   & \textbf{65704}                     & 59381           \\
                    & CLASS\_C & 5702        & 3039  & 4601   & {\ul 8270}    & 8117                               & \textbf{8466}   \\
                    & CLASS\_D & 3359        & 1729  & 1936   & {\ul 6492}    & 6404                               & \textbf{6572}   \\
                    & CLASS\_E & 7210        & 3210  & 4955   & 8397          & {\ul 8503}                         & \textbf{9363}   \\ \hline
\multirow{5}{*}{38} & CLASS\_A & {\ul 53451} & 26346 & 45360  & 50973         & 51056                              & \textbf{64194}  \\
                    & CLASS\_B & 19949       & 9538  & 20343  & {\ul 30556}   & \textbf{31084}                     & 30385           \\
                    & CLASS\_C & 2790        & 1536  & 2800   & 3269          & {\ul 3328}                         & \textbf{3666}   \\
                    & CLASS\_D & 1883        & 1118  & 1526   & \textbf{3439} & {\ul 3385}                         & 3148            \\
                    & CLASS\_E & 4090        & 2105  & 3005   & {\ul 4664}    & 4640                               & \textbf{5356}   \\ \hline
\multicolumn{2}{c}{Average}    & 40086       & 16564 & 24865  & 48859         & {\ul 49040}                        & \textbf{50972}  \\ 
\hline
\hline
\end{tabular}
}
\end{table}

\subsection{Analysis of Steganalysis Resistance}\label{sec6f}
Steganalysis resistance plays a crucial role in evaluating steganography algorithm, as it determines whether the steganography algorithm provides adequate security. In this section, we first evaluated the steganalysis resistance performance of the six comparison schemes at different payloads using various intra and inter steganalysis methods including Zhao \cite{28}, Sheng \cite{29}, Li \cite{30}, Huang \cite{31}, Zhai\cite{32} and Dai\cite{33}. Although these steganalysis algorithms are not specifically aimed at detecting video steganography based on block structure, they are effective in some cases. Therefore, we employ them to comprehensively analyze and compare the anti-steganalysis performance of our schemes. The experimental results are shown in TABLE \ref{tab5}. In the table, the symbol $(\rightarrow 50)$ represents a steganalysis detection rate close to 50\%, indicating better resistance to steganalysis. As can be seen in the table, all video steganography schemes based on block structure exhibit strong resistance to both intra and inter steganalysis methods except for Tew \cite{23} in resisting Sheng \cite{29}. The reason is that the six steganalysis algorithms are not specifically designed for video steganography based on block structure. Although Sheng\cite{29} is a steganalysis algorithms based on intra prediction modes, but it still extracts the features of detecting the CU partition difference before and after recompression. Therefore, it can detect Tew\cite{23}, since Tew\cite{23} greatly changes the CU block structure. 

\begin{table}[htpb]
\caption{
The detection accuracy $(\rightarrow 50\%)$ of intra and inter steganalysis of the six schemes\label{tab5}}
\centering
%\resizebox{\linewidth}{!}{
\scriptsize
% 2. 减小列间距（默认 6pt，改为 3pt 或更小）
\setlength{\tabcolsep}{6pt}
% 3. 减小行间距（默认 1，改为 0.8 或 0.7）
\renewcommand{\arraystretch}{0.8}
\begin{tabular}{@{}ccccc@{}}
\toprule
\multirow{2}{*}{Steganalysis}                    & \multirow{2}{*}{Steganography} & \multicolumn{3}{c}{Payload}                      \\ \cline{3-5} 
                                                 &                                & 0.1 bpc        & 0.3 bpc        & 0.5 bpc        \\ \hline
\multirow{6}{*}{Zhao\cite{28}}  & Tew\cite{23}  & 54.78          & 61.22          & 65.57          \\
                                                 & Dong\cite{24} & 51.48          & 53.04          & 54.43          \\
                                                 & Yang\cite{25} &  50.43    & {\ul 52.00}    & {\ul 52.52}    \\
                                                 & Wang\cite{26} & 51.83          & 52.35          & 53.04          \\
                                                 & Proposed ($8\times8$) & {\ul50.35}          & 52.66          &  53.79\\
                                                 & Proposed                       & \textbf{49.96} & \textbf{50.43} & \textbf{51.13} \\ \hline
\multirow{6}{*}{Sheng\cite{29}} & Tew\cite{23}  & 72.87          & 77.22          & 86.78          \\
                                                 & Dong\cite{24} & 54.26          & 61.74          & 66.61          \\
                                                 & Yang\cite{25} & \textbf{51.13} & 61.22          & 63.65          \\
                                                 & Wang\cite{26} & 53.22          & {\ul 58.43}    & \textbf{60.35} \\
                                                 & Proposed ($8\times8$) & {\ul52.35}          & 58.89          &  62.19\\
                                                 & Proposed            &  52.89    & \textbf{57.57} & {\ul 61.91}    \\ \hline
\multirow{6}{*}{Li\cite{30}}    & Tew\cite{23}  & 48.35          & 49.34          & 50.24          \\
                                                 & Dong\cite{24} & 49.04          & {\ul 49.68}    & {\ul 50.05}    \\
                                                 & Yang\cite{25} & {\ul49.45} & 49.51          & 50.21          \\
                                                 & Wang\cite{26} & 48.70          & 49.45          & 49.74          \\
                                                 & Proposed ($8\times8$) & \textbf{49.55}          & 50.43          &  50.45\\
                                                 & Proposed           &  49.34    & \textbf{49.86} & \textbf{49.97} \\ \hline
\multirow{6}{*}{Huang\cite{31}} & Tew\cite{23}  & 48.75          & 49.22          & 50.14          \\
                                                 & Dong\cite{24} & {\ul 49.37}    & 49.45          & 50.26          \\
                                                 & Yang\cite{25} & \textbf{49.91}          & {\ul 49.68}    & 49.80          \\
                                                 & Wang\cite{26} & 48.17          & 49.39          & {\ul 50.09}    \\
                                                 & Proposed ($8\times8$) & 48.45          & 49.35          &  49.88\\
                                                 & Proposed                       & 49.28 & \textbf{49.97} & \textbf{50.03} \\ \hline
\multirow{6}{*}{Zhai\cite{32}}  & Tew\cite{23}  & 48.58          & 50.61          & 51.65          \\
                                                 & Dong\cite{24} & \textbf{49.57}          & 50.78          & 51.48          \\
                                                 & Yang\cite{25} & {\ul 49.39}    & {\ul 50.30}    & 51.83          \\
                                                 & Wang\cite{26} & 48.35          & 50.43          & {\ul 51.13}    \\
                                                 & Proposed ($8\times8$) & 48.55          & 50.58          &  51.45\\  
                                                 & Proposed                       & 49.22 & \textbf{49.74} & \textbf{50.09} \\ \hline
\multirow{6}{*}{Dai\cite{33}}   & Tew\cite{23}  & 51.79          & 52.55          & 52.67          \\
                                                 & Dong\cite{24} & 51.50          & 52.37          & 53.00          \\
                                                 & Yang\cite{25} & 51.61          &  52.00    &  52.33    \\
                                                 & Wang\cite{26} & {\ul 51.37}    & 52.19          & 53.50          \\
                                                 & Proposed ($8\times8$) & 51.52          & {\ul51.43}          & {\ul 52.13}\\
                                                 & Proposed                       & \textbf{51.00} & \textbf{51.33} & \textbf{52.01} \\ \hline
\end{tabular}
%}
\end{table}

In order to more accurately assess the steganalysis resistance performance of the six comparison schemes, we further use the CBSSM proposed in Section \ref{sec3b} to extract the detection features in the stego video from the six comparison schemes at different payloads (referred to as CBSSM steganalysis in the following), and then a binary classification is performed using a LibSVM classifier with an RBF kernel. The dataset is generated using all 32 test sequences listed in TABLE \ref{tab1}, which includes both original and stego videos. The dataset is randomly split into a training set and a test set at a 1:1 ratio. we repeat the above process 100 times and take the average to obtain the detection results. 

As shown in TABLE \ref{tab6}, as the payload increases, the CBSSM steganalysis resistance of all algorithms declines. Similarly, a higher QP slightly reduces steganalysis resistance, as it amplifies the recompressed stego video’s deviation from the original, causing more noticeable block structure changes and easier detection. Under different $QPs$ and payloads, both schemes we proposed achieve optimal and suboptimal performance. Notably, when $QP=26$ and payload=0.1 bpc, the detection accuracy of our schemes is close to 50\%, indicating excellent CBSSM steganalysis resistance. This is because the modification of each CU is limited to a depth of one and our three-level distortion function applies larger distortion to blocks that remain unchanged before and after recompression, aiming to preserve their block structure as much as possible, while applying smaller distortion to blocks that change significantly before and after recompression, thus changing the blocks that undergo significant alterations. The slightly higher detection accuracy of Proposed ($8\times8$) can be attributed to the fact that it exclusively modifies the features of $8\times8$ blocks, whereas Proposed distributes the embedding-induced alterations across blocks of all sizes.

\begin{table}[htpbt]
\caption{The detection accuracy $(\rightarrow 50\%)$ of CBSSM steganalysis of the six schemes}\label{tab6}
\centering
%\resizebox{\linewidth}{!}{
\scriptsize
% 2. 减小列间距（默认 6pt，改为 3pt 或更小）
\setlength{\tabcolsep}{6pt}
% 3. 减小行间距（默认 1，改为 0.8 或 0.7）
\renewcommand{\arraystretch}{0.8}
\begin{tabular}{ccccc}
\hline
\multirow{2}{*}{$QP$} & \multirow{2}{*}{Steganography} & \multicolumn{3}{c}{Payload}                      \\ \cline{3-5} 
                      &                                & 0.1 bpc        & 0.3 bpc        & 0.5 bpc        \\ \hline
\multirow{6}{*}{26}   & Tew\cite{23}  & 78.09          & 91.13          & 96.32          \\
                      & Dong\cite{24} & 67.83          & 77.91          & 83.30          \\
                      & Yang\cite{25} & 66.43          & 81.22          & 85.26          \\
                      & Wang\cite{26} &  63.65    &  69.57    &  72.87    \\
                      & Proposed ($8\times8$) & {\ul52.96}          & {\ul58.35}          &  \textbf{62.11}\\
                      & Proposed                       & \textbf{50.61} & \textbf{56.35} & {\ul63.65} \\ \hline
\multirow{6}{*}{32}   & Tew\cite{23}  & 85.91          & 91.45          & 97.54          \\
                      & Dong\cite{24} & 68.17          & 81.56          & 88.52          \\
                      & Yang\cite{25} &  67.23    & 82.61          & 88.87          \\
                      & Wang\cite{26} & 68.35          &  74.96    &  85.74    \\
                      & Proposed ($8\times8$) & {\ul64.76}          & {\ul68.59}          &  {\ul75.62}\\
                      & Proposed                       & \textbf{63.56} & \textbf{66.43} & \textbf{72.87} \\ \hline
\multirow{6}{*}{38}   & Tew\cite{23}  & 86.97          & 93.19          & 98.97          \\
                      & Dong\cite{24} & 68.35          &  82.36    & 92.70          \\
                      & Yang\cite{25} &  70.78    & 82.91          & 91.29          \\
                      & Wang\cite{26} & 71.48          & 84.17          &  88.14    \\
                      & Proposed ($8\times8$) & \textbf{64.58}          & {\ul75.29}          &  \textbf{76.53}\\
                      & Proposed                       & {\ul65.04} & \textbf{74.61} & {\ul78.61} \\ \hline
\end{tabular}
%}
\end{table}

In contrast, the detection accuracy for the other four algorithms remains above 60\%. Even at a payload of 0.5 bpc, both of our proposed algorithms maintain a detection accuracy below 80\%, while the others almost exceed 90\%. In summary, CBSSM steganalysis effectively detects existing block structure steganography algorithms and our algorithm shows much stronger resistance. Specifically, Tew \cite{23} modifies CU size to $8\times8$, which significantly disrupts the original CU block structure and makes it susceptible to detection by CBSSM features. For Dong \cite{24}, SCEDM model cannot keep structure similarity of CUs, which leads to insufficient resistance to CBSSM feature. For Yang \cite{25}, the multi-stage STC embedding leads to significant modifications. For Wang \cite{26}, only modifying the $8\times8$ CU resulted in an abnormality in $8\times8$ CBSSM feature, making it easy to be detected.

\subsection{Ablation Study}\label{sec6g}
To further demonstrate the effectiveness of the proposed three-level distortion function, we conducted ablation experiments, in which the environment is the same as described in Section \ref{sec6a}, and the anti-steganalysis, $\Delta PSNR$, and $BIR$ with and without distortion function are shown in TABLE \ref{tab7}. 

\newcommand{\tabincell}[2]{\begin{tabular}{@{}#1@{}}#2\end{tabular}}%表格文字换行使用的代码
\begin{table}[htpb]
\caption{The results of the proposed scheme with and without the distortion function}\label{tab7}
\centering
\begin{tabular}{ccccc}
\hline
\tabincell{c}{Mapping \\Rule}             & \tabincell{c}{Three-level \\Distortion}   & \tabincell{c}{CBSSM \\Steganalysis (\%)}    & $\Delta PSNR$ (dB)   & $BIR$            \\ \hline
\checkmark & $\times$              & {\ul77.04}    & {\ul 0.0416}    & {\ul 0.0190}                  \\
\checkmark & \checkmark & \textbf{66.43} & \textbf{0.0386} & \textbf{0.0177}               \\ \hline
\end{tabular}
\end{table}

It can be seen from the table that by introducing the distortion function, the performance in resisting CBSSM steganalysis has been significantly improved. Because our three-level distortion function prioritizes the embedding of secret information in CUs whose block structure changes after recompression, it can effectively prevent feature anomalies caused by block restoration phenomenon. Meanwhile, the $\Delta PSNR$ and $BIR$ of the proposed algorithm are also reduced when employing the distortion function. This is because the distortion function takes into account the RDO value, thereby minimizes embedding distortion and effectively mitigates the increase in $BIR$.

\begin{figure}[h]
    \centering
    \includegraphics[width=0.9\linewidth, trim=0cm 0cm 0cm 0cm, clip]{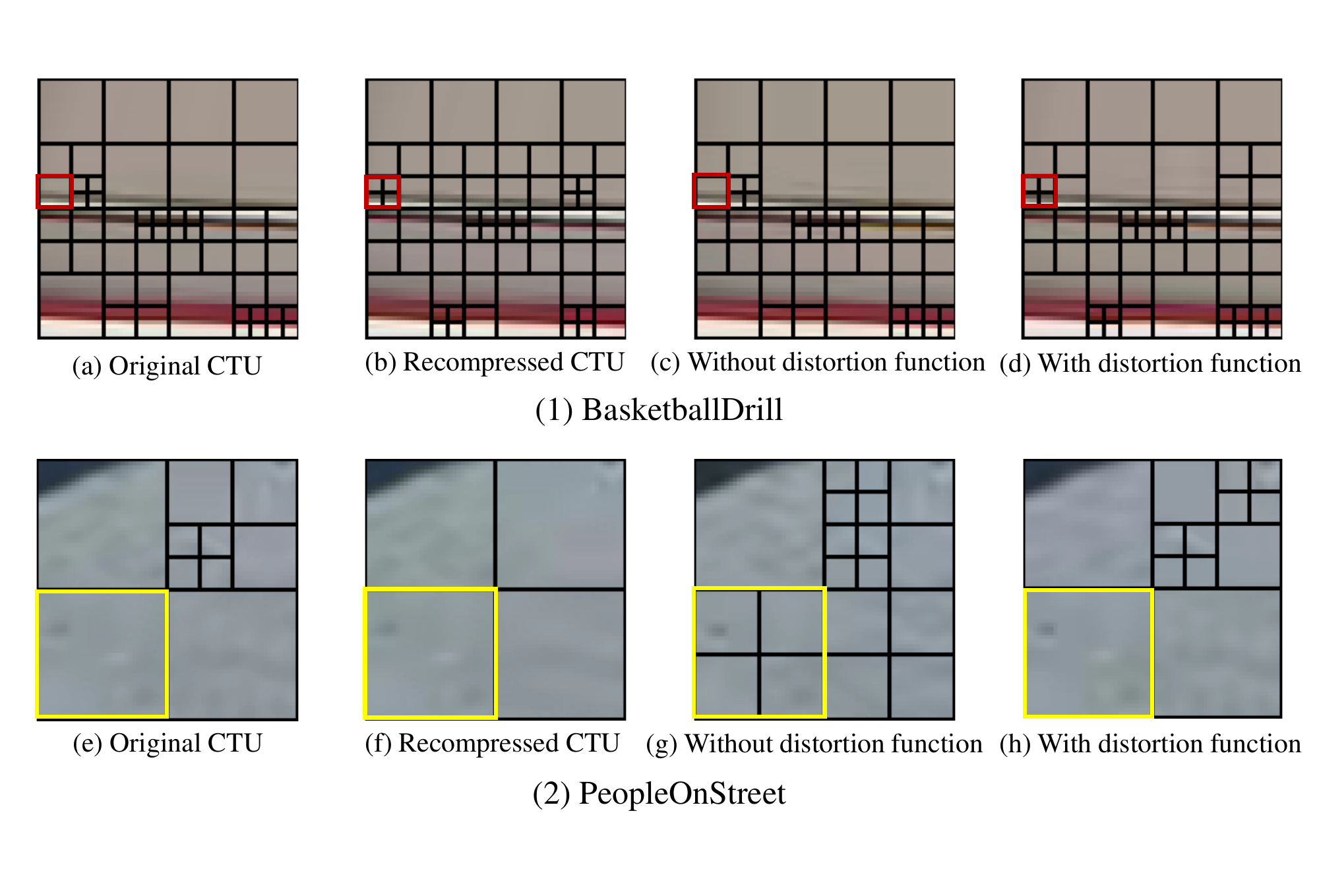}    %左下右上
    \captionsetup{justification=justified, singlelinecheck=false}
    \caption{The example CTUs of the proposed scheme with and without the distortion function}
    \label{fig.ablation}
\end{figure}

To further verify the effectiveness of the proposed three-level distortion, we compare the block structure of CTUs after steganography with and without the distortion. Fig. \ref{fig.ablation} shows two examples which is obtained from \#1 frame of ``BasketballDrill" and ``PeopleOnStreet" sequence at $QP=32$ and payload=0.5 bpc.

In ``BasketballDrill" sequence, the CU marked by red exhibits a changed block structure after recompression. Without three-level distortion, the block structure after steganography remains consistent with the original. However, when three-level distortion is applied, the block structure changes, indicating that this distortion enables secret information to be embedded in CUs whose structure changes after recompression. Conversely, in ``PeopleOnStreet" sequence, the CU marked by yellow retains its block structure after recompression. Without three-level distortion, the block structure after steganography changes. However, with the three-level distortion, the structure remains unchanged. This demonstrates that three-level distortion effectively avoids embedding secret information in CUs whose block structures remain stable after recompression.

%(这里解释不清楚，实际上我们的失真是针对隐写分析的，为什么这里是PSNR和BIR)

\section{Conclusion}\label{sec7}
This paper proposes a H.265/HEVC video steganography based on multiple CU size and block structure distortion. We first explain the block structure restoration phenomenon, leveraging the phenomenon, a CU block structure stability metric which contains block quantity unchanged metric and block structure invariance metric are designed to successfully reveal the reason for the insufficient anti-steganalysis of existing video steganography algorithms based on block structure. Afterwards, we propose the video steganography scheme with a mapping rule and a three-level distortion function to securely embed secret information in carriers. Experimental results show that the CU block structure stability metric is highly sensitive to changes in block structure. Meanwhile, the proposed video steganography scheme maintains the block structure while ensuring the video quality, embedding capacity, anti-steganalysis and bitrate performance. Our future work will focus on further improving anti-steganalysis performance under high embedding capacity.

%参考文献
\bibliographystyle{ieeetr}
\bibliography{ref}

\end{document}